\documentclass{article}
\begin{document}

\begin{center}
{\large \bf Time invariance violation in photon-atom \\ 
and photon-molecule interactions.}
\end{center}

\vspace{3 pt}
\centerline{\sl V.A.Kuz'menko}
\vspace{5 pt}
\centerline{\small \it Troitsk Institute of Innovation and Fusion Research,}
\centerline{\small \it Troitsk, Moscow region, 142092, Russian Federation.}
\vspace{5 pt}
\begin{abstract}

	A direct experimental proof of strong T-invariance violation in
 interactions of the photons with atoms and molecules exists in the molecular
 physics. 

\vspace{5 pt}
\end{abstract}

\vspace{5 pt}

\hspace{\textwidth minus 10cm}\parbox{9cm}{\it "Even for the physicist [a] 
description in plain language will be a criterion of the degree of 
understanding that has been reached". 
Werner Heisenberg, "Physics and Philosophy" 
(Harper Bros., New York, p.168, 1958).}	

\vspace{12 pt}

	The searches for the P- and T- invariance violation are being 
conducted in the atomic physics for many years [1]. As concerns the 
T-invariance violation, these searchers are based on the indirect 
indications, namely, they assume the neutron, electron and atoms to possess 
an electric dipole momentum [2]. Time reversibility invariance demands, 
that the probability amplitudes of the direct and reverse processes should 
be equal [3]. The problem of the direct test of T-invariance is not 
discussed in the elementary particle physics. Apparently, this is due to 
the obvious facts, that reverse in time, for example, of the $K^0$ meson 
decay is practically unreal.

	On the contrary, the reverse in time of the process of interaction 
of the photon with atoms and molecules, in principle, can be easily 
implemented experimentally. The time invariance in this case, apparently,
implies the absorption cross-section of the photons to be equal to that of 
their stimulating emission. We can effect on the atoms or molecules by 
laser radiation to excite them. Then we can influence by the laser 
radiation once more to deexcite them. After that, it is possible to measure 
the cross-section of the direct process (the absorption) and that of the 
inverse one (the stimulating emission) and to compare them. There is no 
principal difficulty.

	The main problem of the experimental technique is that the 
absorption lines are usually very narrow, so that the Doppler effect 
hinders to measure their natural widths and cross-sections. In other cases, 
when the natural linewidth happens to be wider then the Doppler width, the 
lifetime of the excited states proves to be very small. It does not allow 
separating in time the processes of excitation and monitoring the excited 
states.

	 However, in the molecular physics there is a unique object, which is 
characterized by an unusual combination of properties. On the one hand, it 
has very large homogeneous spectral width of optical transition (several 
orders of magnitude greater than the Doppler width). On the other hand, it 
has a long lifetime of the excited states compared to that of the spontaneous 
emission ($>1ms$), that allows to separate in time the processes of the laser 
excitation and monitoring. Experiments with this object in late $1980^{th}$ 
had yielded surprising results, which were not given any theoretical 
explanation, the works with this object were stopped and for more than ten 
years nobody dared to return to it. Those experiments directly demonstrated 
T-invariance violation in interactions of the photons with molecules.

	This unique object is the so-called wing of the spectral line or the 
wide component of the line [4]. It appears as a certain wide continuum in the 
absorption spectrum of the medium size molecules (having usually 4--10 atoms).
 Its nature, apparently, is connected with some features of the vibrational 
motion of atoms in the molecules [5]. It is important, that we have several 
reliable experimental proofs of the existence of the wide component. For 
example, from the results of works [6,7], where an excitation of the $SF_6$ 
molecules by radiation of a pulsed $CO_2$-laser in the conditions of a 
molecular jet was studied, it is possible to obtain precisely the absorption 
cross-section of the wide component  $\sigma =6\cdot{10^{-20}} \ cm^2$ and 
to evaluate it's lorentzian width ($\sim150GHz$). Direct observation of the 
wide component as a continuum was conducted in the work [8], where absorption 
of radiation of the continuous wave $CO_2$-laser by the $SF_6$ molecules was 
studied in the conditions of the molecular beam. From the results of this 
work it is possible to evaluate the absorption cross-section of the continuum 
($\sim{10^{-19}}cm^2$). Finally, the lorentzian shape of a continuum was 
confirmed by detection of the far wings of molecular absorption bands, which 
have a lorentzian shape and have intensity, corresponding to the intensity of 
the continuum near the center of the absorption band [4]. 

	Thus, the continuum characterizes mainly the homogeneous absorption 
of molecules. And probing the excited states by radiation of the second 
$CO_2$-laser demonstrates presence in the absorption spectrum of a sharp dip 
with a width   $\sim 450 kHz$ [8]. It is a typical indication of the 
heterogeneity of a spectrum. In late $1980^{th}$ this effect was not given any 
explanation. The only explanation is that the spectral width of the reverse 
optical transition should be $\sim3\cdot{10^5}$ times less, than that of the 
forward transition. Moreover, in fact, this ratio can be even much 
greater [9]. 

	In the same conditions, amplification of the probe laser radiation 
was also observed. If the fluence of the first $CO_2$-laser radiation is 
less then $20 \mu j/cm^2$, about 0,01\% of the total $SF_6$ molecules are 
excited in the beam [8]. Amplification of the probe laser radiation in this 
case is possible only if the cross-section of the reverse transition exceeds 
that of the direct one by a factor of $>10^4$. Thus, these experiments 
clearly show that the direct and the reverse processes differ very much in 
their cross-sections and spectral widths. However, integrated cross-sections 
of the direct and reverse processes (the Einstein coefficients) can be 
identical. Observations of the Rabi oscillations vote for the benefit of 
such an assumption. 

	The experiments, mentioned above, can be considered as a direct proof 
of the T-invariance violation. At the same time, numerous facts, accumulated 
in the atomic and molecular physics, can be regarded as indirect indications 
of this phenomenon. An existing semiclassical theory of the optical 
transitions [10,11] allows satisfactory description of the physical effects 
from the apparent, formal side. But on this basis it is practically 
impossible to get any concrete information on the physical nature of those 
processes, which underlie the observable effects. 

	Perhaps, most brightly this feature manifests itself in explanation 
of the population transfer effect at sweeping the resonance conditions. If 
the laser is turned precisely in a resonance with the optical transition in 
a two-level system, than, according to the theory, the population of levels 
should oscillate with Rabi frequency. If the wavelength of the laser 
radiation changes and passes through a resonance (sweeping of the resonance 
conditions), it can be regarded as decrease in the effective energy of the 
laser pulse. In this case, reduction in the number of the Rabi oscillations 
can be expected. However, only complete and irreversible transfer of 
population from one level on another in one sweeping cycle is observed in the 
experiments. Manipulations with vectors and series in the model of a rotating 
wave allow the theorists to declare that they can explain this effect [12]. 
However, they cannot reveal the physical reason of the incipient asymmetry 
and irreversibility of the population transfer process. The difference in the 
spectral width of the forward and backward optical transitions quite suits 
as such a physical reason. In this case sweeping of the resonance conditions 
must lead in a natural and inevitable way to the complete population transfer.

	A similar situation takes place in many other cases also. 
T-invariance violation, described above, is a good physical basis for 
explanation of  the nature of such effects, as amplification without 
inversion [13], a phase conjugation [14], a photon echo [15], 
an electromagnetically induced transparency [16], a coherent population 
trapping [17], the Autler-Townes effect [18] and others. 

	An existing semiclassical theory of the optical transitions is based 
on the assumption about the complete T-invariance of the process of 
interaction of photons with atoms and molecules. Taking the fact of strong 
T-invariance violation demands rather radical revision of the theory. But 
more important now is to reanimate and to prolong the experiments with the 
line wings [8].

\end{document}